\documentstyle[12pt,aaspp4,natbib209]{article}

\newcommand{\lsim}{\lower.4ex\hbox{$\;\buildrel <\over{\scriptstyle\sim}\;$}}
\newcommand{\gsim}{\lower.4ex\hbox{$\;\buildrel >\over{\scriptstyle\sim}\;$}}

\begin{document}

\title{Search for Planetary Candidates within the OGLE Stars}

\author{Adriana V. R. Silva\altaffilmark{1} and 
Patricia C. Cruz\altaffilmark{1}}

\altaffiltext{1}{CRAAM, Universidade Presbiteriana Mackenzie, Rua da 
Consola\c c\~ao, 896, S\~ao Paulo, SP 01302, Brazil {\it 
(asilva@craam.mackenzie.br)}}

\begin{abstract}

We propose a method to distinguish between planetary and stellar 
companions to stars which present a periodic decrease in 
brightness, interpreted as a transit. Light curves from a total 
of 177 stars from the OGLE project were fitted by the model 
which simulates planetary transits using an opaque disk in front 
of an image of the Sun. The simulation results yield 
the orbital radius in units of stellar 
radii, the orbital inclination angle, and the ratio of the 
planet to the star radii. Combining Kepler's third law with a 
mass-radius relation for main sequence stars, it was possible to 
estimate values for the masses and radii of both the primary and 
secondary objects. This model was successfully tested with the 
confirmed planets orbiting the stars HD 209458, TrES-1, 
OGLE-TR-10, 56, 111, 113, and 132. The method consists of 
selecting as planetary candidates only those objects with 
primary densities between 0.7 and 2.3 solar densities 
(F, G, and K stars) and secondaries with 
radius less than 1.5 Jupiter radius. The method is not able to 
distinguish between a planet and a dwarf star with mass less 
than 0.1 $M_\odot$, such as OGLE-TR-122. We propose a selection 
of 28 planetary candidates (OGLE-TR-49, 51, 55, 63, 71, 76, 90, 
97, 100, 109, 114, 127, 130, 131, 134, 138, 140, 146, 151, 155, 
159, 164, 165, 169, 170, 171, 172, and 174) for high resolution 
spectroscopy follow up.
\end{abstract}

\section{Introduction}

Since the discovery of the first extra-solar planet orbiting a
solar-like star \citep{mayor1995}, there has been an increased 
interest in the search for planets around other stars. 
Presently, over 150 planets have already been discovered (see 
The Extra-Solar Planets Encyclopedia at 
www.obspm.fr/encycl/encycl.html). The only way to decide if a 
companion object is a planet or a dim star is through 
the measurement of its mass.

There are basically two methods for planet detection commonly 
applied, both of which are heavily biased toward giant planets 
in close in orbits. The most used method is radial velocity 
measurements of the star motion around the center-of-mass of the
planet-star system. This method, however, yields only the mass 
lower limit,
$M_p sin(i)$, where $M_p$ is the secondary mass and $i$ is the
inclination angle of the orbit. 
Another way to detect planets is through the observation of 
eclipses caused by planetary transits in front of the parent 
star.  Because of the limited precision attained with ground based
observations, only objects with sizes of the order or 
larger than Jupiter, 
which cause $\sim 1-2$\% dips in the light curve, can be
detected. Due to their similar sizes, however, the transiting 
objects could be extra-solar planets, brown dwarfs, or M-type 
dwarfs. Moreover, shallow transits due to blending may also be 
mistaken as planetary eclipses. Both methods are complementary, in the 
sense that it is possible to fully characterize the companion by 
its mass and radius, plus the orbital parameters (semi major 
axis, inclination angle, and period). 

Until 2002, the transit of only one planet, HD 209458b, had been 
observed \citep{charbo2000, henry2000}. 
This scenario changed when the Optical Gravitational Lensing
Experiment (OGLE) found evidences of transits in the light-curves
of 177 stars
\citep{udalski2002a, udalski2002b, udalski2002c, udalski2003,
udalski2004}. The presence of a planet was confirmed on five
of those stars by follow up of radial velocity measurements: 
OGLE-TR-10 \citep[also suggested by][]{dreizler2002}, OGLE-TR-56
\citep{konacki2003b, konacki2003a, torres2004}, OGLE-TR-111 
\citep{pont2004}, OGLE-TR-113 \citep{konacki2004, bouchy2004} 
and OGLE-TR-132 \citep{bouchy2004}. The TrES multisite 
transiting planet survey also yields the detection of a 
transiting Jupiter-sized planet \citep{alonso2004}.

Previous authors have also investigated the nature of the OGLE 
star companions to identify stellar binaries. The methods used varied 
from the detection of ellipsoidal variation \citep{drake2003, sirko2003}, 
to spectroscopic
surveys \citep{torres2004, konacki2003a, dreizler2002}, and 
infrared spectroscopy \citep{gallardo2005}. \citet{drake2003}
searched for sinusoidal modulations in the light curves due 
to the ellipsoidal form of the stellar envelope caused by the 
proximity of a second star. Extending the analysis of 
\citet{drake2003}, \citet{sirko2003} found ellipsoidal 
variations above the 3 sigma level in 30 of the studied stars, 
indicating the presence of a red dwarf companion. 
\citet{tingley2005} have also devised a method which employs an 
exoplanet diagnostic $\eta$ in order to identify the best 
planetary candidates from a transit search. 

Here we propose a test for distinguishing between planetary and 
stellar companions based on transit observations, and apply this 
method to the stars observed by the OGLE survey. The next section 
describes the transit simulation using an image of the Sun. 
Section~\ref{test} presents the result of the best fit to the OGLE 
data along with the discussion. Finally, the conclusions are presented in the 
last section. 

\section{The method: ``planetary" transit across the Sun}\label{model}

A white-light image of the Sun from Big Bear Solar Observatory 
is used to simulate the parent star \citep[as in][]{silva2003}. 
The advantage of using an image of the Sun is that limb 
darkening is already taken into account. It is true that some 
stars have quadratic limb darkening \citep[e.g. HD 
209458,][]{brown2001} instead of the linear limb darkening, as 
in the solar case, however, due to the large uncertainty 
of the OGLE data it is not possible to distinguish between these 
two. 

The secondary object is represented by an opaque disk of
radius $R_p/R_s$, where $R_p$ is the ``planet" radius whereas 
$R_s$ is the radius or the primary star. The position of 
the ``planet" in its orbit is calculated for each time interval 
according to the orbital parameters: inclination angle, $i$, and 
semi-major axis, $a/R_s$ (in units of the stellar radius). All 
simulations were performed assuming the orbit to be circular, 
that is null eccentricity. This assumption is probably valid due 
to the strong gravitational tidal forces and the close proximity 
to the primary star. The orbital periods used were those given 
by the OGLE project. 

Figure~\ref{par} shows the effect of varying the three 
parameters used in the transit simulation, namely, $R_p/R_s$, 
$a/R_s$, and $i$. The effect of increasing the secondary radius, 
$R_p/R_s$, is to strengthen the dimming in the light curve 
(Figure~\ref{par}a). By decreasing the distance of the secondary 
to its parent star, $a/R_s$, the phase interval of the transit 
is increased (Figure~\ref{par}b). The latitude of the transit is 
determined by the orbit inclination angle and the orbital 
radius, which in turn changes both the phase interval of the 
transit and its depth (Figure~\ref{par}c).

The light curve of a particular star containing a dimming, 
interpreted as 
a transit from a secondary faint object, is then fit by the 
simulation. The parameters of the transit are chosen from the 
best fit to the data, that is, minimum
$\chi^2$. 

In order to test the method, observations of known planets such 
as HD209458b \citep{deeg2001} and TrES-1 \citep{alonso2004} were 
used. The best transit fit to the data (crosses) is shown as a 
solid line in Figure~\ref{tran}, 
and the parameters obtained are given in Table~\ref{tres}. 
The same method was applied to the stars observed by the OGLE
project which, through radial velocity measurements, had their 
planetary companions verified, that is, OGLE-TR-10, 56, 111, 
113, and 132.  For each object, the first line of the table 
shows the parameters obtained from the observations, as listed 
in The Extra-Solar Planets Encyclopedia. The parameters obtained from 
the best fit, listed on Table~\ref{tres} (second line for each 
object, denominated ``model"), agree with the published values 
within the uncertainty estimates. In order to estimate the 
uncertainty of the model parameters ($R_p/R_s$, $a/R_s$, and 
$i$), a 1000 light curves were generated by varying randomly
each of the 3 parameters within reasonable intervals. From 
these, we selected only those light curves which values remained 
within the envelope defined by 
$\pm 1\sigma$ of the observed light curve. This standard
deviation was calculated from the data outside the transit.
Finally, the uncertainty was estimated from the distribution 
of the parameter values of these selected light curves.

For a second test to the model, we used the data 
from OGLE-TR-122, which has been
identified as binary system \citep{pont2005} with masses $M_1=0.98 M_\odot$ 
and $M_2=0.092 M_\odot$, the latter low mass star has a radius of
only $R_2=0.12 R_\odot$. The results from our model (also shown in 
Figure~\ref{tran} and listed on Table~\ref{tres}) are $M_1=0.81 M_\odot$, 
$M_2=0.05 M_\odot$, and $R_2=0.094 R_\odot$, agreeing quite well with
the observations. Yet another test performed was to fit the model to
a synthetic lightcurve which was generated for a binary system with
a primary star of 4 $M_\odot$ and a secondary of 0.32 $M_\odot$ with
radius $R_2=3.92 R_J$, on an orbit of $i=84^\circ$ and semi-diameter
of 0.075 A.U.. After adding random noise to the light curve (shown
in Figure~\ref{tran}), the model yields $M_1=3.75 M_\odot$, 
$M_2=0.29 M_\odot$, $R_2=3.62 R_J$, 
$i=85.3^\circ$, and orbital radius of 0.074 A.U.. As can be seen,
once more the model seems to reproduce well the expected values.

\section{Planetary versus stellar companion}\label{test}

From a total of 177 stars detected by the OGLE 
project \citep{udalski2002a, udalski2002b, udalski2002c, udalski2003} whose light curve  presented periodic dimming, four were discarded for not 
having their period listed (OGLE-TR-43 to 46). A remainder of 173 
stars were analyzed, and the secondary object and its orbital parameters obtained by minimizing the difference between the data
and the transit simulation, as described in the previous section.
The parameters, $R_p/R_s$, $a/R_s$, and $i$, obtained from 
the model best fit to the data are 
plotted as histograms in Figure~\ref{hfit}, for the 173 stars 
studied. 

The radius of the secondary
object, $R_p/R_s$, in units of the primary radius, are plotted in the top
left panel of Figure~\ref{hfit}. The dashed vertical line 
in this panel represents the relative size of 
Jupiter with respect to the Sun, and indicates that the majority
of companions are larger than the relative size of Jupiter.
The orbital distance, $a/R_s$
determined from the model assuming circular orbits were found 
to be between 2-20 stellar radii, and
are shown in the top right panel of Figure~\ref{hfit}.
Most orbital inclination 
angles are close to $90^\circ$, which is expected for the detection 
of transits. Angles smaller than $80^\circ$ may be an indication
of grazing eclipses and their respective results should not be considered. The bottom right panel of the figure shows
the density of the primary star inferred from the observations, estimated by the following relationship obtained 
using Kepler's third law:

\begin{equation}
\rho={M_1+M_2 \over R_1^3} = {4\pi^2 \over G P^2} \left({a \over R_s}\right)^3
\end{equation}

\noindent where $P$ is the period and $a/R_s$ is the orbital
radius, both quantities directly inferred from the fit without
any further assumptions.
If $M_2<<M_1$ then $\rho$ is the density of the primary star, when this is 
not true, however, this density will be overestimated. 
Only those stars with densities with
values within the dotted lines, corresponding to F, G, and K stars, will be further considered, as discussed in the next section.

The aim of this work is to propose a method for selecting the best
planetary candidates for a follow up with radial velocities
measurements, which is very expensive due to the time 
requirement on large telescopes with high precision spectrographs.
Below we explain how it is possible to infer the absolute
values of the mass and radius of both the primary and secondary objects, and therefore determine
the companion object radius and its orbital distance in absolute
values.

In order to calculate the four unknowns $M_1$ (or $M_s$), $R_1$ (or $R_s$), $M_2$, and $R_2$, the masses and radii of the primary 
and secondary objects respectively, we need four equations. 
The first one is Kepler's third law: 
 
\begin{equation}
\left({a \over R_s}\right)^3 = {G P^2\over 4 \pi^2 R_1^3} (M_1+M_2),
\end{equation}

\noindent where $a/R_s$ is obtained from the best fit to 
the data. The second equation is given by the ratio of the primary
and secondary radii: $R_p/R_s=R_2/R_1$, where the left handside is determined from the
depth of the transit in the observed lightcurve. The two 
last equations are obtained by applying a mass-radius
relationship for main sequence stars, $R_s/R_\odot = (M_s/M_\odot)^{0.8}$ 
\citep{allen2000}, for both the primary and the secondary
object. Thus one can determine the 
mass and radius of both objects from the four equations listed
above. Here we have considered both objects to be main sequence
stars. Note that for a companion with a tenth of a solar mass, 
this mass-radius relationship yields  a radius of 
about 1.5 times that of Jupiter, making it impossible to
distinguish between a planet and a dwarf star based only on transit data.  

Using the above equations, we calculate the value of 
the primary radius in units of solar radius, and thus
determine the absolute value of the orbital 
semi-diameter in A.U.. Histograms of these parameters
are plotted in Figure~\ref{hpar}. Primary stellar masses
in units of solar mass (plotted on the top left panel of 
Figure~\ref{hpar}) vary between 0.5 and 7 $M_\odot$,
with a maximum value around one solar mass. 
The top right panel of the figure displays the mass of
the secondary object, which is not the real value if
this is a planet. The radius of the secondaries, 
ranging from 0.5 to 8 Jupiter's radius, is
shown in the bottom left panel. The most
common companion radius being about 1.5 $R_J$. 
The semi-diameter of a circular orbit is shown in the 
bottom right panel, with values in 
the range between 0.01 and 0.15 A.U., peaking at 0.05 A.U..

As mentioned above, in the case that the secondary
object is a planet, then the value of $M_2$ will not be real, however, its radius $R_2$
will be a good estimate of the true value. 
If $M_1>>M_2$ then the mass of the primary star is given by

\begin{equation}
{M_s \over M_\odot} = \left({G P^2 M_\odot \over 4 \pi^2 R_\odot^3 (a/R_s)^3} \right)^{1/1.4}
\end{equation}

\noindent and its radius is obtained from the mass-radius
relationship given above. A similar analytical model has 
been proposed by \citet{seager2003}, using the same stellar
mass-radius relation, however, in their model the authors 
do not consider the stellar limb darkening. In the case 
where the mass of the secondary, $M_2$, is much less than 
that of the primary, $M_1$, then in the above equation 
$M_s=M_1$. When the secondary is a dwarf star, 
however, $M_s=(M_1^{2.4}/(M_1+M_2))^{1/1.4}$, and the resulting
$M_s$ will be less than $M_1$. Hence equation 2 should not 
be used to estimate the mass of the primary star.

To rule out stellar companions, first we need to eliminate
large stars such as A, B, and O stars with radius 
at least 10 times larger than a 0.3-0.5 $M_\odot$ secondary 
star that can cause a transit of 1-2\% dimming. For this 
we use the density of the primary star. Since, here we are
interested in only F, G, and K stars so that the secondary
companion will have radii of the order of Jupiter, we will
consider further only those stars with densities between 0.7 
and 2.3 that of the Sun. The lower limit excludes big stars 
such as A, B, or O stars, whereas the upper density limit was
chosen as that of the densest star known to harbor planets 
(see Table~\ref{tres}). Results of densities larger than 
about 2 either represent M dwarf stars or maybe binary systems.
By adopting this criterion, only 72 candidates remained
(including the 5 stars already known to harbor planets listed 
in Table~\ref{tres}), yielding a total of 42\% of the OGLE
stars. 

In order to better constrain the planetary candidates, our initial 
list of 67 OGLE stars (the 5 bonafide planets were excluded)
were checked against the stellar companions confirmed by the 
previous work cited in the Introduction. From these a total of 11 
stars were either listed as showing ellipsoidal variation or are 
giants, implying that their companion is not an exoplanet. 

Therefore, a more conservative criterion was established 
in order 
to improve our method: besides the constraint on the density 
of the primary star, for the secondary to be considered a 
planet candidate its radius has to be less than 1.5 $R_J$. 
Note that due to the adopted mass-radius relationship,
this criterion is equivalent to considering only those 
secondaries with masses less than 0.1 $M_\odot$.
This shortened the candidate list to the 28 stars displayed 
in Table~\ref{tog}, plus OGLE-TR-122. The uncertainties were 
estimated in the same way as described previously.
The parameters of these selected stars 
are shown as the gray histograms in Figure~\ref{hpar}. 

These 28 objects were confronted with the results of
\citet{tingley2005},
where a secondary is considered a planet candidate if 
the parameter $\eta_p<1$. Only 6 stars (OGLE-49, 151, 159, 
165, 169, and 170) failed the comparison having $\eta$ 
larger than unity. 
From high resolution spectroscopy, \citet{pont2005} showed that 
OGLE-TR-122 is a very low mass star ($0.092 M_\odot$), with 
about the same size as Jupiter. Since its mass is less than 
one tenth that of the primary, this stellar companion could not 
have been detect by our method. Nevertheless the detection of
such low mass stars are interesting and such objects deserve
further study.

\section{Conclusions}

A method for selecting the best planetary candidates has been 
presented here and applied to 173 stars with evidence of low 
luminosity companion transits observed by the OGLE project. The 
method consists of simulating a transit by an opaque disk, the 
``planet", in front of a white light image of the Sun, representing 
the star. Besides the orbital period which is directly obtained 
from the OGLE data, three parameters are determined from a 
best fit of the model to the data: {\it (i)} ratio of companion
to star radii, $R_p/R_s$; {\it (ii)} orbital semi-diameter (assuming
circular orbit) in units of stellar radius, $a/R_s$; and 
{\it (iii)} orbit inclination angle, $i$. 

In order to obtain absolute values for the secondary radius and 
its orbital distance, it is necessary to determine the primary 
radius. Combining Kepler's 3rd law with a mass-radius 
relationship for main sequence stars ($R_s\propto M_s^{0.8}$) 
and the observed transit depth, it was possible to infer the 
mass and radius of the primary and secondary objects.
This method worked quite well for the 7 known planetary 
companions: HD209458b, TrES-1, OGLE-TR-10, 56, 111, 113, and 
132, as can be seen from Table~\ref{tres}.

At first only stars with densities between 0.7 and 2.3 solar
density (corresponding to F, G, and K stars) were considered.
Excluding the 5 known planets, a total of 67 stars resulted. 
When comparing these stars with those studied by previous 
authors, we noted that this was not a sufficient criterion 
for ruling out stellar companions. Hence, a further constraint 
on the radius of the secondary was adopted: only those companions
with radius $<1.5 R_J$ were considered, similar to 
\citet{dreizler2002}. We point out that this method is not 
able to distinguish between true planets and dwarf stars with 
masses of the order or less than 0.1 $M_\odot$, since their 
sizes are similar to that of Jupiter (e.g. OGLE-TR-122). 
Nevertheless, the study of these low mass stars is also 
interesting.

These criteria resulted in 28 planet candidates, listed in 
Table~\ref{tog}, comprising
less than 16\% of the OGLE stars. We propose that these stars 
should have high precision spectroscopic follow up in order to 
confirm or not, by the estimate of their mass, if they truly are 
planets.

\acknowledgments

PCC acknowledges support from the Brazilian agency FAPESP (grant 
number 03/04541-4). We are gratefunl to Dr. A. Udalski and his 
team for making the OGLE data available in the Internet, and to 
the anonymous referee for the suggestions which improved the paper. 
AVRS thanks Dr. A.J.R. da Silva for fruitfull discussions.

\bibliographystyle{apj}

\bibliography{apj-jour,ref}

\newpage

\begin{table}[h]
\caption{Best fits from the model}
\begin{tabular}{clllll}
\hline
Star &  & $M_s$ ($M_\odot$) & $R_p$ ($R_J$) & $a$ (A.U.) & $i$ ($^\circ$) \\
\hline
HD209458 & obs.$^a$ & 1.05 & 1.32$\pm$0.05 & 0.045 & 86.1$\pm$0.1 \\
   & model & 1.14$\pm$0.22 & 1.35$\pm$0.20 & 0.047$\pm$0.003 & 87.2$\pm$0.6 \\
TrES-1   & obs.$^a$ & 0.87$\pm$0.03 & 1.08$\pm$0.18 & 0.0393$\pm$ 0.0007 & 88.2$\pm$1.0 \\
   & model & 0.86$\pm$ 0.21 & 1.14$\pm$ 0.25 & 0.0390$\pm$ 0.0028 & 88.4$\pm$1.2 \\
OGLE-TR-10 & obs.$^a$ & 1.22$\pm$0.04 & 1.24$\pm$0.09 & 0.0416$\pm$0.007 & 89.2$\pm$2.0 \\
   & model & 1.09$\pm$0.20 & 1.32$\pm$0.18 & 0.0428$\pm$0.0026 & 88.1$\pm$0.6 \\
OGLE-TR-56 & obs.$^a$ & 1.04$\pm$0.05 & 1.23$\pm$0.16 & 0.0225$\pm$0.0004 & 81.0$\pm$2.2 \\
   & model & 0.80$\pm$0.23 & 0.86$\pm$0.19 & 0.0206$\pm$0.0020 & 85.4$\pm$0.6 \\
OGLE-TR-111 & obs.$^a$ & 0.82$\pm$0.15 & 1.00$\pm$0.13 & 0.047$\pm$0.001 & 86.5-90 \\
   & model & 0.96$\pm$0.21 & 1.16$\pm$0.19 & 0.049$\pm$0.003 & 88.1$\pm$0.6\\
OGLE-TR-113 & obs.$^a$ & 0.77$\pm$0.06 & 1.08$\pm$0.07 & 0.0228$\pm$0.0006 & \\
   & model & 0.72$\pm$0.18 & 1.09$\pm$0.21 & 0.0223$\pm$0.0019 & 90.0$\pm$0.3  \\
OGLE-TR-132 & obs.$^a$ & 1.35$\pm$0.06 & 1.13$\pm$0.08 & 0.0306$\pm$0.0008& 85$\pm$1\\
   & model & 1.2$\pm$0.3 & 0.90$\pm$0.17 & 0.0292$\pm$0.0025 & 90.0$\pm$0.3  \\
\hline
OGLE-TR-122 & obs.$^b$ & 0.98$\pm$0.14 & 1.17$\pm$0.16 & & 88-90 \\
   & model & 0.81$\pm$0.07 & 0.96$\pm$0.08 & 0.0672$\pm$0.0021 & 89.3 $\pm$0.5 \\
\hline
\end{tabular}
$^a$ The observational parameters are taken from The Extra-solar
 Planets Encyclopedia (www.obspm.fr/encycl/encycl.html).

$^b$ \citet{pont2005}
\label{tres}
\end{table}

\begin{table}
\caption{Planetary candidates}
\begin{tabular}{cccccr}
\hline
OGLE-TR & $M_s$ ($M_\odot$) & $R_p$ ($R_J$) & $a$ (A.U.) & $i$ ($^\circ$) \\
\hline  
  49 & 0.99$\pm$0.20 & 1.43$\pm$0.24 & 0.0377$\pm$0.0027 & 89.9 $\pm$0.3 \\
  51 & 0.79$\pm$0.13 & 1.22$\pm$0.16 & 0.0263$\pm$0.0014 & 89.9 $\pm$0.3 \\
  55 & 1.30$\pm$0.34 & 1.36$\pm$0.28 & 0.0462$\pm$0.0039 & 85.9 $\pm$0.6 \\
  63 & 1.21$\pm$0.35 & 1.07$\pm$0.24 & 0.0217$\pm$0.0022 & 87.6 $\pm$0.6 \\
  71 & 0.87$\pm$0.13 & 1.28$\pm$0.17 & 0.0485$\pm$0.0024 & 87.1 $\pm$0.5 \\
  76 & 1.06$\pm$0.24 & 1.28$\pm$0.22 & 0.0330$\pm$0.0024 & 86.9 $\pm$0.6 \\
  90 & 0.84$\pm$0.19 & 1.19$\pm$0.21 & 0.0189$\pm$0.0014 & 82.3 $\pm$0.6 \\
  97 & 1.44$\pm$0.42 & 1.43$\pm$0.33 & 0.0151$\pm$0.0016 & 79.5 $\pm$0.6 \\
 100 & 0.99$\pm$0.35 & 1.21$\pm$0.33 & 0.0172$\pm$0.0019 & 89.6 $\pm$0.4 \\
 109 & 1.62$\pm$0.31 & 1.18$\pm$0.16 & 0.0161$\pm$0.0011 & 89.0 $\pm$0.6 \\
 114 & 0.90$\pm$0.14 & 1.16$\pm$0.14 & 0.0270$\pm$0.0014 & 87.8 $\pm$0.6 \\
 127 & 0.83$\pm$0.23 & 0.85$\pm$0.18 & 0.0285$\pm$0.0026 & 84.3 $\pm$0.6 \\
 130 & 0.76$\pm$0.15 & 1.33$\pm$0.21 & 0.0509$\pm$0.0032 & 86.5 $\pm$0.4 \\
 131 & 0.82$\pm$0.31 & 0.75$\pm$0.22 & 0.0278$\pm$0.0032 & 85.4 $\pm$0.6 \\
 134 & 1.31$\pm$0.16 & 1.17$\pm$0.11 & 0.0587$\pm$0.0025 & 88.3 $\pm$0.6 \\
 138 & 0.77$\pm$0.23 & 0.71$\pm$0.15 & 0.0343$\pm$0.0031 & 87.7 $\pm$1.0 \\
 140 & 0.85$\pm$0.25 & 1.06$\pm$0.27 & 0.0418$\pm$0.0041 & 87.0 $\pm$1.2 \\
 146 & 0.80$\pm$0.15 & 0.94$\pm$0.14 & 0.0373$\pm$0.0024 & 88.2 $\pm$1.1 \\ 
 151 & 0.97$\pm$0.36 & 1.02$\pm$0.25 & 0.0252$\pm$0.0027 & 89.2 $\pm$0.8 \\
 155 & 1.20$\pm$0.38 & 0.95$\pm$0.23 & 0.0630$\pm$0.0059 & 87.8 $\pm$1.1 \\
 159 & 1.16$\pm$0.25 & 1.24$\pm$0.19 & 0.0339$\pm$0.0024 & 88.5 $\pm$1.0 \\
 164 & 1.40$\pm$0.36 & 1.09$\pm$0.22 & 0.0422$\pm$0.0038 & 87.1 $\pm$1.2 \\
 165 & 1.07$\pm$0.28 & 1.02$\pm$0.20 & 0.0405$\pm$0.0035 & 87.6 $\pm$1.1 \\
 169 & 1.16$\pm$0.44 & 0.80$\pm$0.22 & 0.0405$\pm$0.0047 & 88.6 $\pm$1.0 \\
 170 & 1.43$\pm$0.33 & 1.07$\pm$0.17 & 0.0567$\pm$0.0042 & 88.8 $\pm$0.9 \\
 171 & 0.74$\pm$0.26 & 0.61$\pm$0.14 & 0.0290$\pm$0.0031 & 87.8 $\pm$1.2 \\
 172 & 1.24$\pm$0.61 & 0.86$\pm$0.31 & 0.0310$\pm$0.0050 & 88.5 $\pm$1.1 \\
 174 & 0.81$\pm$0.21 & 0.58$\pm$0.13 & 0.0389$\pm$0.0038 & 87.5 $\pm$1.1 \\
\hline
\end{tabular}
\label{tog}
\end{table}

\newpage

\figcaption{Influence of varying the parameters of the simulation:
a) secondary radius, $R_p/R_s$; b) orbital radius, $a/R_s$; and c)
inclination angle, $i$. 
\label{par}}

\figcaption{Best fit for the transit model (solid line) for stars HD209458, TrES-1, OGLE-TR-10, 56, 111, 113, 132, and 122. The parameters obtained are
listed on Table~\ref{tres}. The crosses represent the data. Also shown in the
top right panel is the modeled binary system used as a test to the method.
\label{tran}}

\figcaption{Histograms of the parameters obtained from the best fit to 
the transit data: secondary radius,
$R_p/R_s$, orbital radius $a/R_s$, inclination angle $i$, and primary
star density (in units of solar density) for the 173 OGLE stars.
\label{hfit}}

\figcaption{Histograms of the secondary parameters after using the mass-radius relationship for main sequence stars: primary and secondary masses in units of 
solar mass, secondary radius, $R_p$ in units of Jupiter's radius, and orbital
semi-diameter $a$ in A.U. for the 173 OGLE stars. The shaded histograms 
correspond to the selected stars (including those known to harbor planets)
which have densities between 0.7 and 2.3 solar density and secondary objects
smaller than 1.5 Jupiter radius.
\label{hpar}}

\newpage
\plotone{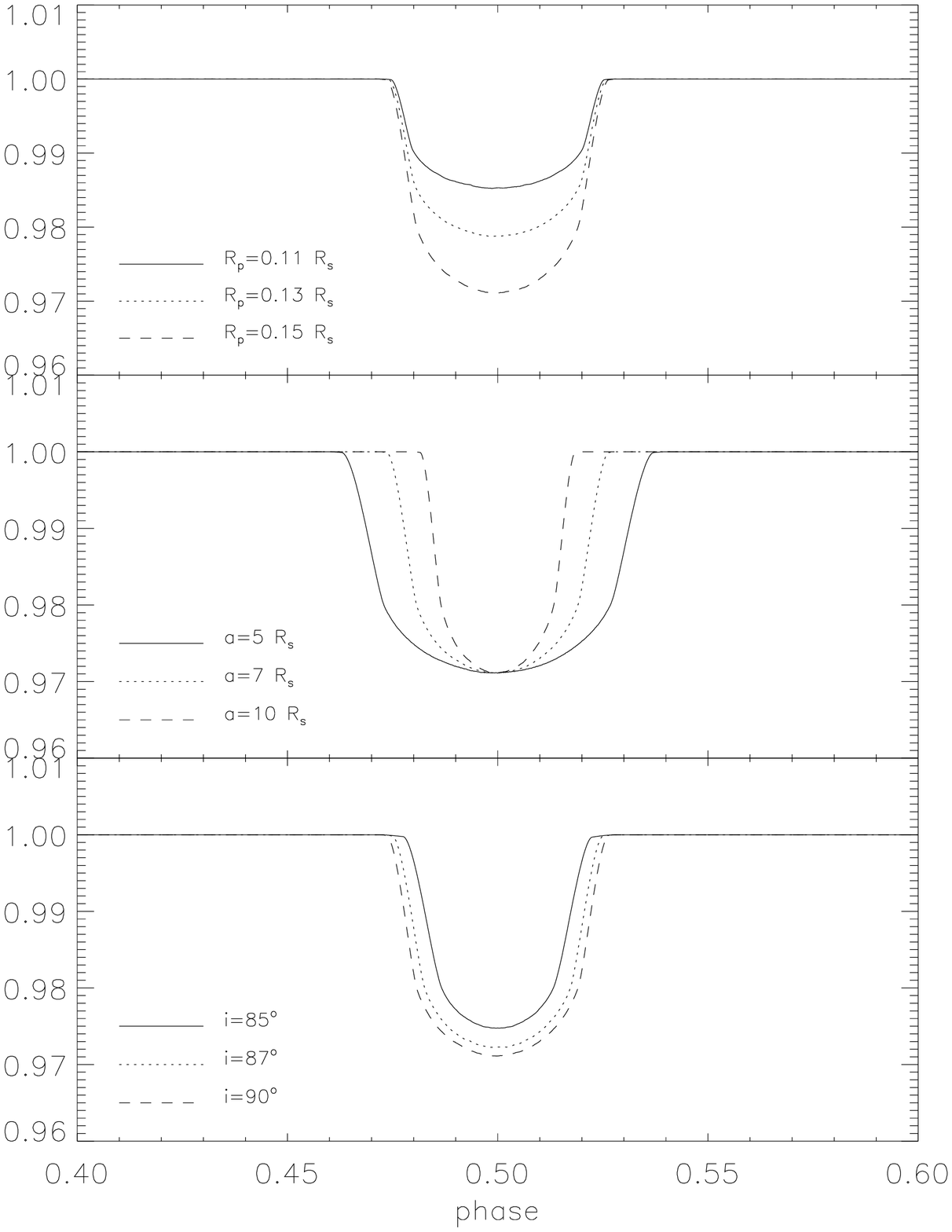}
\newpage
\plotone{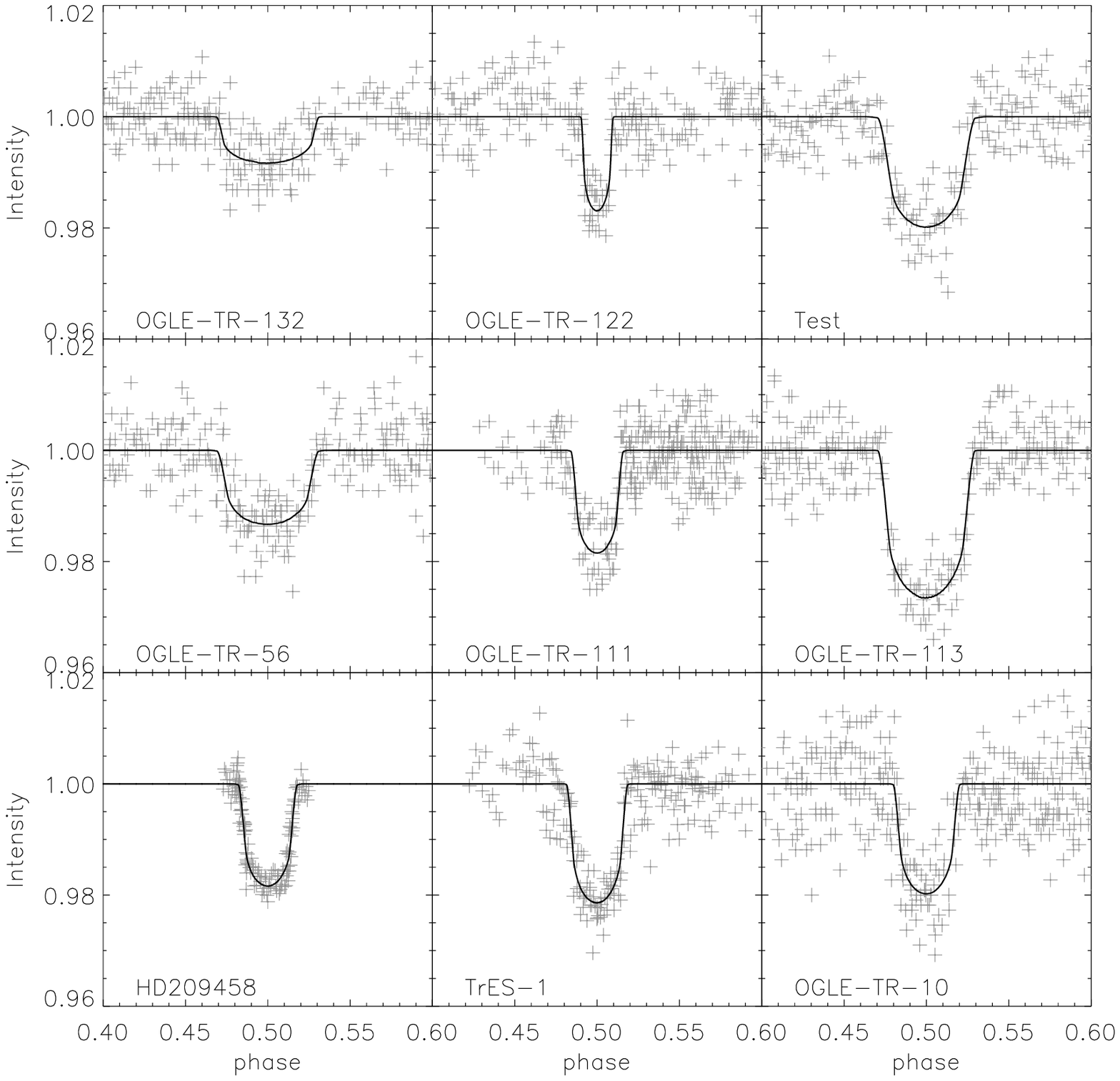}
\newpage
\plotone{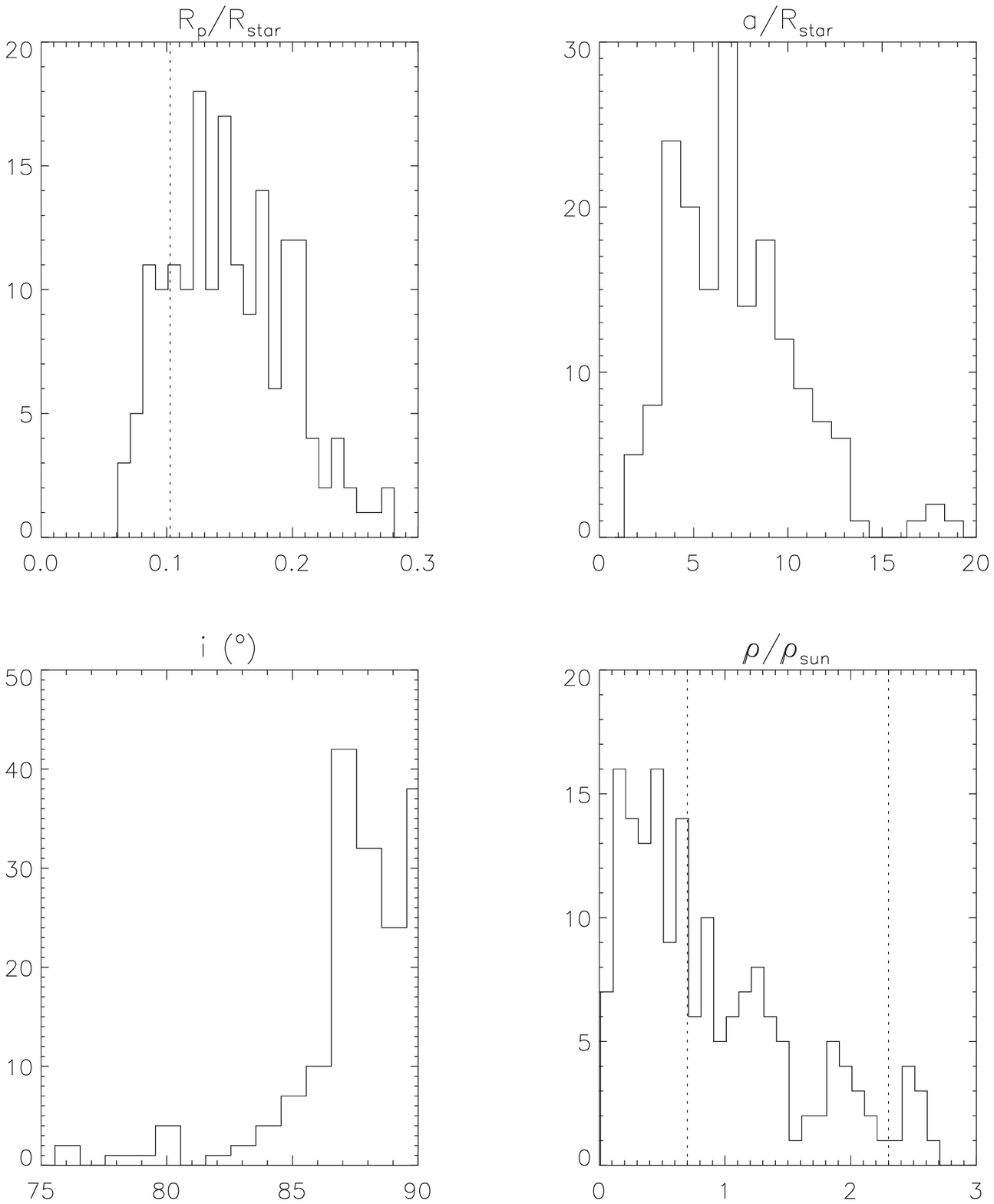}
\newpage
\plotone{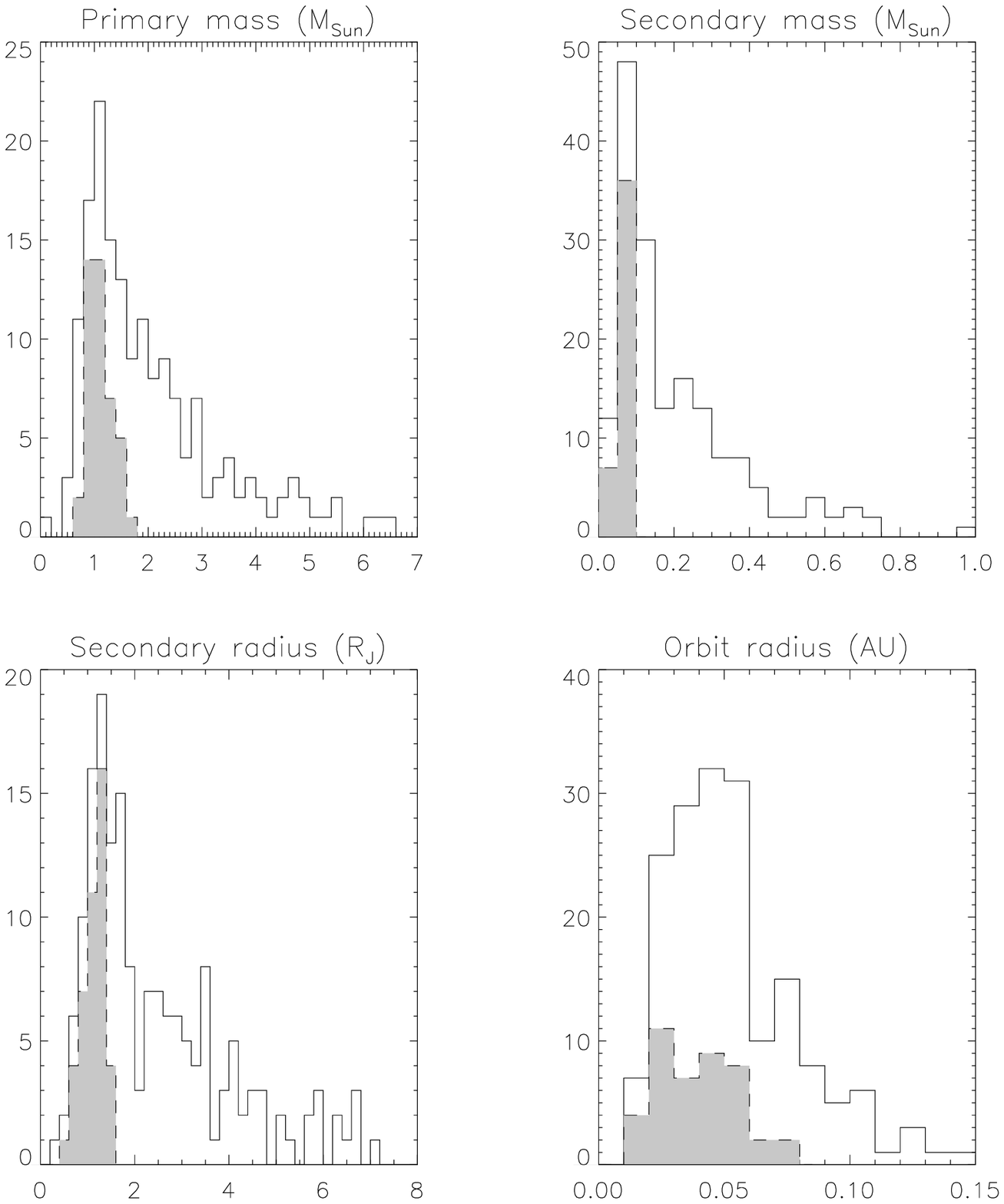}

\end{document}